\documentclass[prb,aps,floatfix,amsmath,twocolumn]{revtex4}

\usepackage{amsmath}
\usepackage{epsfig}
\usepackage{array}
\usepackage{verbatim}
\usepackage{hyperref}
\usepackage{amssymb}
\usepackage{multirow}
\usepackage{graphicx}
\usepackage[export]{adjustbox}

\usepackage{color}
\usepackage{ulem}

\usepackage[cspex,bbgreekl]{mathbbol}

\newcommand{\re}[1]{(\ref{#1})}

\newcommand{\beg}{\begin{equation}}

\newcommand{\en}{\end{equation}}

\newcommand{\eps}{\varepsilon}
\newcommand{\lam}{\lambda}

\newcommand{\eref}[1]{Eq.~(\ref{#1})}
\newcommand{\esref}[1]{Eqs.~(\ref{#1})}

\renewcommand{\Re}{\mathrm{Re}}
\renewcommand{\Im}{\mathrm{Im}}
\newcommand{\Tr}{\mathrm{Tr}\,}

\renewcommand{\emph}{\textit}

\newcommand{\beq}{\begin{equation}}

\newcommand{\eeq}{\end{equation}}
\newcommand{\barray}{\begin{eqnarray}}
\newcommand{\earray}{\end{eqnarray}}

\usepackage{braket}

\begin{document}

\title{Rotationally invariant ensembles of integrable matrices}

\author{Emil A. Yuzbashyan$^1$, B. Sriram Shastry$^2$, Jasen A. Scaramazza$^1$}
\date{\today}							

\affiliation{$^1$Center for Materials Theory, Department of Physics and Astronomy, Rutgers University, Piscataway, New Jersey 08854, USA\\
$^2$Physics Department, University of California, Santa Cruz, California 95064, USA}

\begin{abstract}
We construct ensembles of \textit{random integrable matrices} with any prescribed number of  nontrivial integrals and formulate  \textit{integrable matrix theory} (IMT) -- a counterpart of random matrix theory (RMT) for quantum integrable models. A type-$M$ family of integrable matrices  consists of exactly $N-M$ independent commuting $N\times N$ matrices  linear in a real parameter. We first develop a rotationally invariant parametrization of such matrices, previously only constructed in a preferred basis. For example, an arbitrary choice of  a vector and two commuting Hermitian matrices defines a type-1 family and vice versa. Higher types similarly involve a random vector and two matrices. The basis-independent formulation allows us to derive the joint probability density for integrable matrices, similar to the construction of Gaussian ensembles in the RMT.
\end{abstract}
 
\date{\today}
\maketitle

\section{Introduction}
\label{intro}

It is well established that random matrix theory (RMT)  describes the universal features of  energy spectra of various  quantum systems\cite{dyson,mehta,PFor,jbohigas,been,jGuhr}. RMT does not, however, capture the  typical behavior observed in exactly solvable many-body models, such as e.g. Poisson level statistics \cite{jpoilblanc,jrabson,jberry,jrelano,jstockmann,jellegaard,jputtner}. Though there exist matrix ensembles (e.g. band matrices\cite{band1,band2}, or an invariant ensemble related to the thermodynamics of non-interacting fermions \cite{Neub}) that display this kind of behavior, it is desirable to have a formulation that is both (i) basis-independent and (ii) stems from a well-defined notion of quantum integrability. The purpose of the present work is an explicit construction of ensembles that have both these properties, thereby bridging the gap and providing the missing ensemble -- integrable matrix theory (IMT) -- for the analysis of quantum integrability.

We recently proposed a simple notion of an integrable matrix (quantum integrability)   that leads to an explicit construction of various classes of parameter-dependent commuting matrices\cite{yuzbashyan,shastry,owusu,owusu1,yuzbashyan1}. In this approach, we consider $N\times N$ Hermitian matrices $H(u)=T+uV$ linear in a real parameter $u$. We call $H(u)$ integrable if it has at least one nontrivial (other than a linear combination of itself and the identity matrix) commuting partner  of the form $\bar{H}(u)=\bar{T}+u\bar{V}$, i.e. $[H(u),\bar{H}(u)]=0$ for all $u$. To appreciate the motivation behind this definition, consider exactly solvable many-body models such as the 1D Hubbard\cite{lieb2,bill,hubbook}, XXZ spin chain\cite{YangYang1,YangYang2,mtaka,ywang} or Gaudin magnets\cite{GaudBook} in the presence of an external magnetic field\cite{SkyExt,Cambia,Gaud2}. Suppose we specialize to a particular number of sites and fix all quantum numbers corresponding to parameter-independent symmetries (e.g. number of spin up and down electrons, total momentum etc. in the case of the Hubbard model). Such blocks are integrable matrices under our definition. Indeed, they are linear in a real parameter (Hubbard $U$, anisotropy,  the magnetic field) and  all have at least one nontrivial integral of motion linear in the parameter. The Gaudin model has as many linear integrals as spins\cite{SkyExt}, while the Hubbard and XXZ models in general have at least one such nontrivial linear integral in addition to more with polynomial parametric dependence\cite{grabXXZ,sirkXXZ,grosHub,shasHub}.

Remarkably, it turns out that merely requiring the existence of commuting partners with fixed parameter-dependence leads to a range of profound consequences. First, it implies a categorization of integrable matrices according to the number of their integrals of motion. We say that $H(u)$ belongs to a type-$M$ \textit{integrable family} if there are exactly $n=N-M$ linearly independent $N\times N$ Hermitian matrices\cite{note1} $H^i(u)=T^i+uV^i$ that commute with $H(u)$ and among themselves at all $u$ and have no common $u$-independent symmetry\cite{note2}, i.e. no $\Omega\ne c\mathbb{1}$ such that $[\Omega, H^i(u)]=0$ for all $i$ and $u$. A type-$M$ family is therefore an $n$-dimensional vector space, where $H^i(u)$ provide a basis, the general member of the family being $H(u)=\sum_i d_i H^i(u)$, where $d_i$ are real numbers. The maximum possible value of $n$ is $n=N-1$ (type-1 or maximally commuting Hamiltonians), while a generic $H(u)$ (e.g. with randomly generated $T$ and $V$) defines a trivial integrable family where $n=1$.

Let us briefly recount further consequences of the commutation requirement and related developments.  Integrable  $3\times3$ matrices first appear in Ref.~\onlinecite{yuzbashyan}. Shastry constructed a class of $N\times N$ commuting matrices\cite{shastry} in 2005, which are type-1 in the above classification. Owusu et. al.\cite{owusu} subsequently developed a transparent parametrization of type-1, an exact solution for their energy spectra, proposed the above notion of an integrable matrix, and proved that  energy levels of any type-1 matrix cross at least once as functions of $u$. Later work parametrized\cite{owusu1} all type-2, 3 and a subclass of type-$M$ for any $M>3$. Let us also note the Yang-Baxter formulation\cite{yuzbashyan1} and eigenstate localization properties\cite{disorder} for type-1.

\begin{figure}
\includegraphics[width=\linewidth]{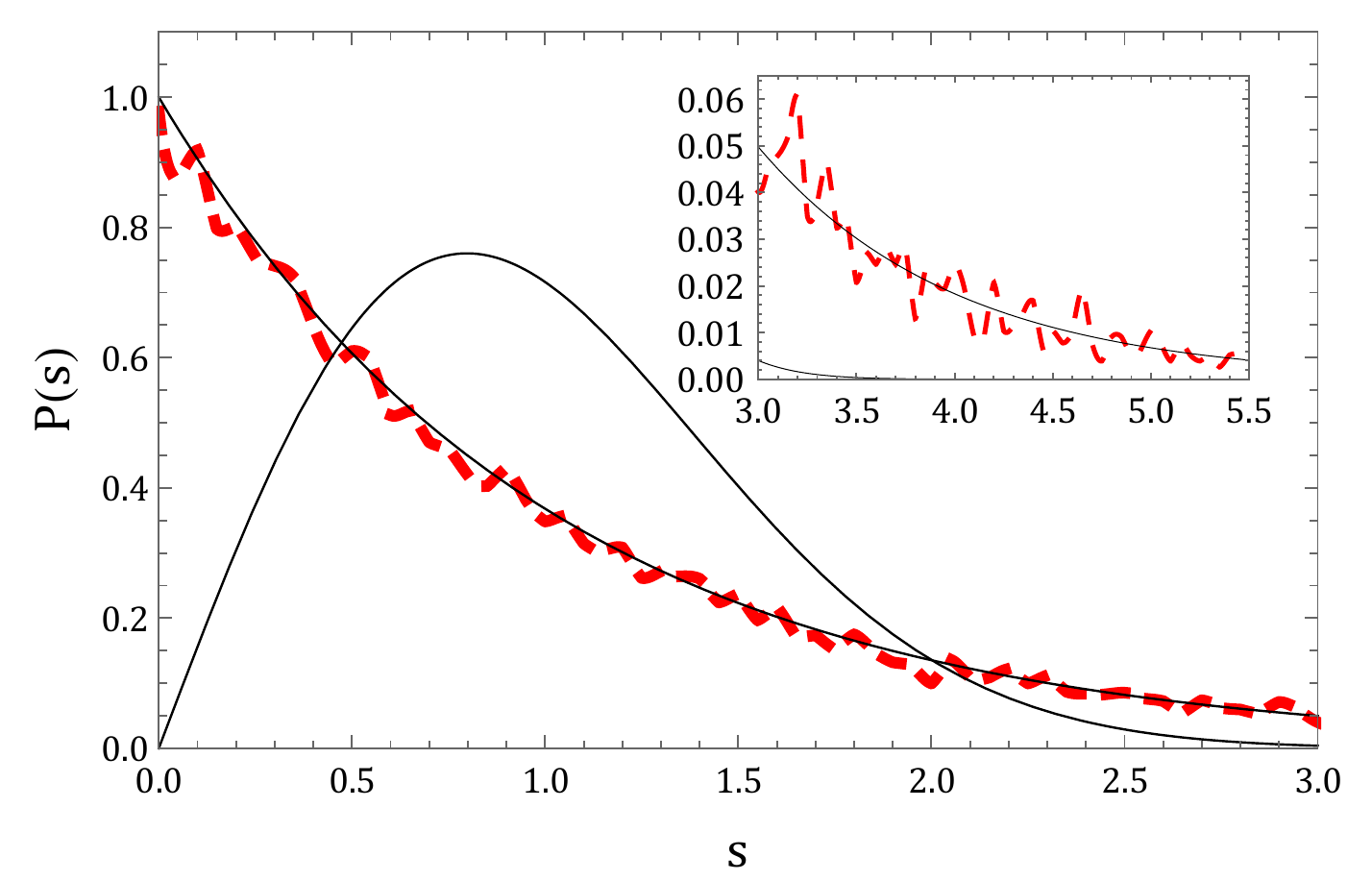}
\caption{(color online) The normalized level spacing distribution $P(s)$ of a single $20000\times 20000$ real symmetric integrable matrix $H(u) = T + uV$ at $u=1$. This matrix, whose construction is detailed in Sect.~\ref{higher}, has exactly 297 nontrivial commuting partners (conservation laws) linear in the parameter $u$ and is therefore type-19703 by our classification. The solid curves are a Poisson distribution $P(s)=e^{-s}$ and the Wigner surmise for real symmetric random matrices $P(s) = \frac{\pi}{2}s\,e^{-\frac{\pi}{4}s^2}$. Poisson level statistics, as shown here, are typical for the invariant integrable matrices  described in this work. Inset: Tails of the same curves.}
\label{statsex}
\end{figure}

However, existing parametrizations are tied to a particular basis, which prevents an unbiased choice of an integrable matrix and obscures the origin of the parameters.   Recall that the invariance of the probability distribution with respect to a change of basis is  a key requirement in  RMT\cite{mehta}. Similarly, a rotationally invariant formulation  is necessary for a proper construction of integrable matrix \textit{ensembles}. Here we first derive such a formulation and then obtain an appropriate probability distribution of \textit{random integrable} matrices with a given number of integrals of motion. In a follow-up work\cite{scaramazza} we will study  level statistics of these ensembles as well as spectral statistics of individual integrable matrices, see Fig.~\ref{statsex} for an example.

More specifically, consider type-1 matrices in the parametrization of Ref.~\onlinecite{owusu}. Up to an arbitrary shift by the identity matrix, a general real symmetric type-1 matrix $H(u)=T+uV$ reads
\beg
H(u)=\frac{1}{2}\sum_{k\ne j} \frac{d_k-d_j}{\eps_k-\eps_j} (\gamma_k\gamma_j p_{kj}-\gamma_j^2 p_k-\gamma_k^2 p_j)+u\sum_{k=1}^N d_k p_k,
\label{type1}
\en
where $d_k, \eps_k, \gamma_k$ are $3N$ arbitrary real numbers, $p_{kj}=|k\rangle\langle j|+|j\rangle\langle k|$,  $p_k=|k\rangle\langle k|$, and $|k\rangle$ are the normalized eigenstates
of $V$ (shared by all $V^i$). This expression immediately yields $kj$-th matrix element of $H(u)$ in the basis where $V$ is diagonal. Parameters $\eps_k$ and $\gamma_k$ specify the commuting family, while $d_k$ pick a particular matrix within the family. Note that $H(u) = \sum_k d_k H^k(u)$, i.e. $H^k(u) = \frac{\partial}{\partial d_k} H(u)$ where $[H^j(u),H^k(u)]=0$, $\forall j,k$. The question is, what is the natural  choice of  $d_k, \eps_k, \gamma_k$? More precisely, what is the probability distribution function of these parameters? For example, we can take $\eps_k$ to be uncorrelated random numbers or eigenvalues of a random matrix from the Gaussian unitary, orthogonal or symplectic ensembles (GUE, GOE, or GSE). Moreover, it turns out that certain choices drastically affect the level statistics, e.g. those where $d_k$ and $\eps_k$ are correlated\cite{yuzbashyan1,scaramazza}.

We will see below that each type-1 family is uniquely specified by a choice of a Hermitian matrix $E$ and a vector $|\gamma\rangle$, $\eps_k$ and $\gamma_k$ in \eref{type1} being the eigenvalues of $E$ and components of $|\gamma\rangle$, respectively. On the same grounds as in RMT, an appropriate choice is therefore to take $E$ from the GUE (GOE for real symmetric, GSE for Hermitian quaternion-real matrices\cite{mehta}) and $|\gamma\rangle$ to be an appropriate random vector.  Note that this choice follows from either rotational invariance of the distribution function combined with statistical independence of the matrix elements or, alternatively, from maximizing the entropy of the distribution\cite{mehta}. Finally, $d_k$ are the eigenvalues of $V$ and we will show that they are distributed as GUE (GOE, GSE) eigenvalues uncorrelated with $\eps_k$. Our construction of integrable matrix ensembles for higher types ($M > 1$) is restricted to the real symmetric case, is more complex and involves the deformation of an auxiliary type-1 family. However, it ultimately amounts to the same choice of $|\gamma\rangle$ and two matrices from the GOE.

\section{Rotationally invariant construction of type-1 integrable matrix ensembles}
\label{type1r}

We start with certain preliminary considerations valid for all types.   The defining commutation requirement, $[H^i(u), H^j(u)]=0$ for all $u$, reduces to three $u$-independent relations
\beg
[V^i, V^j]=0, \quad [T^i, V^j]=[T^j, V^i], \quad [T^i, T^j]=0.
\label{relations}
\en
The second of these relations is equivalent to
\beg
T^i=W^i+[V^i, S], \quad [V^i,W^i]=0,
\label{s}
\en
where $S$ is an antihermitian matrix characteristic of the commuting (integrable) family. Note that $S$ is independent of the element in the family, i.e. for any $H(u)=T+uV$ in the family, $T$ and $V$ are related through 
\beg
T=W_V+[V, S], \quad [V,W_V]=0,
\label{s22}
\en
 with the same $S$. 

Now we specialize to type-1.  Since all $T^i$ commute, they share the same eigenstates $|\alpha_k\rangle$ and therefore
\beg
T^i=\sum_{k=1}^N t^i_k |\alpha_k\rangle\langle\alpha_k|.
\label{decompose}
\en
By definition of type-1, there are $N-1$ linearly independent $T^i$. Together with $\mathbb{1}=\sum_k  |\alpha_k\rangle\langle \alpha_k|$, we have $N$  independent linear equations for $N$ unknown projectors $ |\alpha_k\rangle\langle\alpha_k|$ with a unique solution in terms of $T^i$ for each  $ | \alpha_k\rangle\langle\alpha_k|$. Let $|\alpha_1\rangle\equiv |\gamma\rangle$ for notational convenience. Thus,
\beg
 |\gamma \rangle\langle\gamma|=a_0\mathbb{1}+\sum_i a_i T^i,
\en
where $a_i$ are real numbers (real scalars in the quaternion case). 

Consider an element of the commuting family $\Lambda(u)=a_0\mathbb{1}+\sum_i a_i H^i(u)$. By construction
\beg
\Lambda(u)=|\gamma \rangle\langle\gamma|+uE,
\label{introLam}
\en
where $E$ is an $N\times N$ Hermitian matrix with either complex, real, or quaternion real entries. Moreover, $E$ is nondegenerate, for any degeneracies\cite{degens} in $E$ imply a $u$-independent symmetry $\Omega$  (see Appendix~\ref{appendix1}) contrary to the above definition of an integrable family. Every type-1 integrable family thus contains such a $\Lambda(u)$ given by \eref{introLam} with a rank one $T$-part\cite{rankone}.  We will now show that the converse is also true. In other words, \textit{any} $\Lambda(u)$ (i.e. an arbitrary choice of a vector $|\gamma \rangle$ and a nondegenerate Hermitian matrix $E$) uniquely specifies a type-1 family.

We begin with an arbitrary $\Lambda(u)=\ket{\gamma}\bra{\gamma}+uE$ from which we will construct a type-1 integrable family of matrices $\{H^i(u)\}_{\Lambda}$. We require that $\Lambda(u)$, henceforth known as the ``reduced Hamiltonian'', be an element of this putative family. Then \eref{s} gives
\beg
|\gamma \rangle\langle\gamma| = W_E+[E, S], \quad [E,W_E]=0.
\label{cauchy}
\en
\eref{cauchy} uniquely determines the matrix elements of $S$ as a function of $E$ and $\ket{\gamma}$. We then consider $H(u)=T+uV$ and impose $[\Lambda(u),H(u)]=0$, $\forall u$, which implies  (see \eref{relations} and \eref{s})

\beg
\begin{split}
\quad &[V,E] = 0. \\
\quad&T = W + [V,S], \quad [V,W]=0,\\
\quad &[T,\ket{\gamma}\bra{\gamma}]=0.
\end{split}
\label{relationsL}
\en
The third equation implies  $\ket{\gamma}$ is an eigenstate of $T$. Via a non-essential shift of $T$ by a multiple of the identity we  set the  corresponding eigenvalue to zero, i.e. $T\ket{\gamma}=0$. We will see that the choice of  $V$   in \eref{relationsL} uniquely specifies $T$, and therefore determines $H(u)$. As $E$ is nondegenerate, $\Lambda(u)$ has no permanent degeneracies (eigenvalues degenerate at all $u$) and therefore any $H^i(u)$ and $H^j(u)$ so constructed will satisfy $[H^i(u),H^j(u)]=0$, $\forall u$.

We have thus constructed a type-1 integrable family  $\{H^i(u)\}_{\Lambda}$ from an arbitrary reduced Hamiltonian $\Lambda(u)=\ket{\gamma}\bra{\gamma}+uE$. But from the considerations at the beginning of this section, we know that \textit{all} type-1 families contain a reduced matrix $\Lambda(u)$. It follows that our basis-independent construction, i.e. Eqs.~(\ref{cauchy}-\ref{relationsL}), produces all type-1 matrices.

It is not immediately obvious from Eqs.~(\ref{cauchy}-\ref{relationsL}) that a simple parametrization of matrix elements follows. It is therefore helpful to select a preferred basis and write them in components to demonstrate the feasibility of the construction. In the shared diagonal basis of the matrices $E$ and $V$, \eref{cauchy} implies

\beg
S_{ij} = \frac{\gamma_i \gamma^*_j}{\eps_i-\eps_j},
\label{cauchyB}
\en
where $E = \textrm{diag}(\eps_1,\eps_2, \dots, \eps_N)$ and $\gamma_i$ are the components of $\ket{\gamma}$. The components $\gamma_j$ are either complex, real, or quaternion real, corresponding to the three possibilites for the Hermitian matrix $E$. Therefore $\gamma_j^*$ denotes complex conjugation in the first two cases and quaternion conjugation in the third case. Let $V=\textrm{diag}(d_1,d_2,\dots,d_N)$, then \eref{relationsL} gives

\beg
\begin{split}
T_{ij} = H_{ij}(u) &=  \gamma_i \gamma^*_j \frac{d_i-d_j}{\eps_i-\eps_j},\quad i\ne j,\\
T_{ii} + uV_{ii}  = H_{ii}(u) &= u\,d_i-\sum_{j\ne i}|\gamma_j|^{2}\frac{d_i-d_j}{\eps_i-\eps_j}.
\end{split}
\label{Hcomp}
\en

To determine the common eigenvectors  of $H(u)$, consider the eigenvalue equation $\Lambda(u)|\varphi\rangle =u\lambda|\varphi\rangle$ for the reduced Hamiltonian, 
\beg
|\gamma\rangle\langle\gamma|\varphi\rangle+uE|\varphi\rangle=u\lambda|\varphi\rangle,
\en
where we introduced a factor of $u$ for convenience. In components this yields
\beg
\varphi_k= \frac{\gamma_k}{u(\lambda-\eps_k)}\langle\gamma|\varphi\rangle.
\label{phik}
\en
The ``self-consistency'' condition $\sum_k\gamma_k^*\varphi_k=\langle\gamma|\varphi\rangle$
then implies an equation for $\lambda$
 \beg
\begin{split}
u=\sum_{j=1}^N\frac{|\gamma_j|^2}{\lambda-\eps_j},
\end{split}
\label{lambdaEq}
\en
This equation has $N$ real roots $\lambda_i$ for $i=1,\dots, N$ that play a special role in the exact solution (and the analysis of level crossings) of type-1 Hamiltonians\cite{owusu}.
In particular, the eigenvalues $\eta_i$ of $H(u)$ from \eref{Hcomp} are
\beg
\eta_i = \sum_{j=1}^N\frac{|\gamma_j|^2d_j}{\lambda_i-\eps_j}=\bra{\gamma}V\ket{i},
\label{t1eigenvals}
\en
and the corresponding unnormalized eigenstates $\ket{i}$ according to \eref{phik} read
\beg
\ket{i}_k\equiv\varphi_k^{(i)}=\frac{\gamma_k}{\lambda_i-\eps_k},
 \label{lambdaEq2}
 \en
 Note that these are the components of $\ket{i}\equiv|\varphi^{(i)}\rangle$ in the eigenbasis of $V$ and that $u\lambda_i$ are the eigenvalues of the reduced Hamiltonian.
 
Finally, using Eqs.~(\ref{cauchy}-\ref{relationsL}), one can show that if a family of commuting matrices $H^j(u)$ is Hermitian (real-symmetric, Hermitian quaternion-real) for all $u$,  the corresponding matrices $E$ and $V^j$ are also Hermitian (real-symmetric, Hermitian quaternion-real) and the vector $\ket{\gamma}$ is complex (real, quaternion real) and vice versa.  We will show next in Sect.~\ref{PDF1} that these three choices correspond to selecting these objects from the GUE, GOE or GSE, respectively. Recall that, physically speaking, GUE matrices break time reversal invariance. GOE and GSE matrices are invariant under time reversal, while GSE matrices futhermore break rotational invariance and represent systems with half-integer spin\cite{dyson,mehta}.

\section{Probability density function of type-1 integrable ensemble}
\label{PDF1}

In Sect.~\ref{type1r}, we found that any Hermitian type-1 integrable matrix is specified by the choice of a vector $\ket{\gamma}$ and two Hermitian matrices $E$ and $V$ satisfying $[E,V]=0$. Consider the set of all type-1 $N\times N$ matrices as a random ensemble $\mathcal{H}^1_{N}(u)$ with a probability density function (PDF) $P(\gamma,E,V)$ on the parameters $\ket{\gamma},E$ and $V$. The probability of obtaining a matrix $H(u)\in \mathcal{H}^1_{N}(u)$ characterized by parameters in the region between $(\gamma,E,V)$ and $(\gamma+d\gamma,E+dE,V+dV)$ is $P(\gamma,E,V)\,d\gamma\,dE\,dV$, where
\beg
\begin{split}
d\gamma &= \prod_{i=1}^Nd\,\Re(\gamma_i)\,d\,\Im(\gamma_i),\\
dV &=\prod_{j< i}d\,\Re(V_{ij})\,d\,\Im(V_{ij})\prod_{k}d\,V_{kk}.
\end{split}
\label{measures}
\en
Here we derive a basis-independent $P(\gamma,E,V)$ in a manner similar to the construction of the PDF of the Gaussian RMT ensembles\cite{mehta}. As indicated in \eref{measures}, we will restrict our notation to complex Hermitian matrices. Matrices and vectors with quaternion entries have four real numbers associated to each off-diagonal matrix element and to each vector component. We find that the eigenvalues of $E$ and $V$ (the $\eps_i$ and $d_i$ in \eref{Hcomp}) come from independent GUE, GOE or GSE eigenvalue distributions $\Omega(a)$
\beg
\begin{split}
\Omega(a) \propto \prod_{i<j}|a_i-a_j|^{\beta}e^{-\sum_k a_k^2},
\end{split}
\label{WD}
\en
where $\beta=2$, $1$ and $4$ for the GUE, GOE, and GSE, respectively. The eigenvalue sets are independent essentially because eigenvalues of a random matrix are independent of the eigenvectors, and the $[E,V]=0$ requirement only constrains eigenvectors. The final expression for $P(\gamma,E,V)$ is \eref{RotInvF}, while the corresponding PDF for the parameters from \eref{Hcomp}, denoted $P(\gamma,\eps,d)$, is \eref{RotInvFinal}.

There are two approaches to this derivation, both of which give the same result. First, one can maximize the entropy functional\cite{mehta,Jaynes} $S[P]=-\langle\ln(P)\rangle=-\int_X P(\gamma,E,V)\ln(P(\gamma,E,V))d\gamma\,dE\,dV$ subject to constrained averages, where the set $X$ includes all parameter values such that $|\gamma|^2=1$ and $[E,V]=0$. The constrained averages in this case are $\langle 1 \rangle = 1$, $\langle \Tr E^2 \rangle = \langle \Tr V^2 \rangle = \alpha$, $\alpha \in \mathbb{R}^+$. Alternatively, one may postulate that $(\ket{\gamma},E,V)$ are independent objects, each with its own PDF given by known results from RMT\cite{mehta,PFor} before projecting the product of these PDFs into the constrained space $[E,V]=0$. We use the latter strategy in what follows.

As $\ket{\gamma}$ is independent of $E$ and $V$, we have
\beg
\begin{split}
P(\gamma,E,V) = P(\gamma)\,P(E,V).
\end{split}
\label{RotInv3}
\en
The function $P(\gamma)$ is well known in RMT\cite{PFor}
\beg
\begin{split}
P(\gamma) &\propto \delta \left(1- |\gamma|^2\right),
\end{split}
\label{RotInv5}
\en
which is the only invariant $P(\gamma)$ that preserves the norm $|\gamma|=1$.

We now determine $P(E,V)$, which is the crux of the whole derivation. Consider the PDF $P_0(A,B)$ of two independent $N\times N$ random matrices $A$ and $B$ from the GUE or GOE
\beg
\begin{split}
P_0(A,B)dA\,dB &= P_0(A)P_0(B)dA\,dB,\\
P_0(A) &\propto e^{- \Tr{A^2}}.
\end{split}
\label{IndepEV}
\en
To project $P_0(A,B)$ from \eref{IndepEV} into the constrained space $[A,B] =~ 0$, it is convenient to make a change of variables from the matrix elements $A_{ij}$ (respectively $B_{ij}$) to the eigenvalues $a_i$ ($b_i$) and functions $f$ of eigenvectors $q_i^a$ ($q_i^b$). It is well known that the Jacobian $J(A_{ij};a_i,f(q_i^a))$ of this transformation factorizes\cite{mehta}
\beg
\begin{split}
P_0(A,B)\,dA\,dB &= \Omega(a)\Omega(b)\,da\,db\,df(q^a)\,df(q^b),\\
\\
\Omega(a) &\propto \prod_{j<i}{\left| a_i - a_j \right|^{\beta}}e^{- \Tr{A^2}},\\
da &= \prod_{i} da_i, \quad df(q^a) = \prod_{i}df(q_i^a).
\end{split}
\label{IndepEV1}
\en
We will not specify the precise form of the function $f(q^a)$. Also, by making the change of variables $\{A_{ij}\} \to~ \{a_i,q^a_i\}$, we have implicitly selected a particular gauge of eigenvectors of $A$ (i.e. the eigenvectors have fixed phases). 

If $A$ and $B$ are nondegenerate,  $[A,B]=0$ is equivalent to $q^a_i=q^b_i$, $\forall i$. If $A$ or $B$ have degeneracies, there are many ways for the commutator to vanish, but \eref{IndepEV1} shows $P_0(A,B)$ itself vanishes for any degeneracies. Therefore, the probability $P^{A,B}_{\textrm{comm}}$ that two given matrices $A$ and $B$ commute is
\beg
\begin{split}
P^{A,B}_{\textrm{comm}} = \prod_j\delta\left(f(q_j^a) - f(q_j^b)\right) + \left(\textrm{degen. terms}\right).
\end{split}
\label{IndepEV2}
\en
It follows that the measure $P(E,V)\,dE\,dV$ for commuting matrices $E$ and $V$ is
\beg
\begin{split}
P(E,V)\,dE\,dV &\propto \Omega(\eps)\Omega(v)\prod_j\delta(q_j^{\eps} - q_j^{v})\times \\
&d\eps\,dv\,dq^{\eps}\,df(q^v),
\end{split}
\label{jointcommute}
\en
where $\eps_i$ ($v_i$) are eigenvalues of $E$ ($V$). Thus
\beg
\begin{split}
P(\gamma,E,V)d\gamma\,dE\,dV &\propto \delta \left(1- |\gamma|^2\right) \Omega(\eps)\Omega(v) \times \\
&\prod_j\delta(q_j^{\eps} - q_j^{v})\,d\gamma\,d\eps\,dv\,dq^{\eps}\,df(q^v).
\end{split}
\label{RotInvF}
\en
Now we  integrate out the eigenvectors in order to obtain the joint PDF $P(\gamma,\eps,d)$ for the parameters appearing in \eref{Hcomp}
\beg
\begin{split}
P(\gamma,\eps,d) &\propto \delta \left(1- |\gamma|^2\right)\times\\
&\prod_{i<j}|\eps_i-\eps_j|^{\beta}|d_i-d_j|^{\beta}e^{-\sum_k \eps_k^2}e^{-\sum_k d_k^2},
\end{split}
\label{RotInvFinal}
\en
where we substituted $v_i \to d_i$ in order to be consistent with the notation in previous papers. \eref{RotInvFinal} is particularly significant because it allows one to study the level statistics of the ensemble of $N\times N$ type-1 integrable matrices $\mathcal{H}^1_N$, which according to numerical simulations generally turn out to be Poisson\cite{scaramazza}.

\section{Parameter shifts}
\label{shifting}

Here we consider two parameter shifts that leave the commuting family invariant. The second is useful in the rotationally invariant construction  of
type-$M$ integrable matrices for $M>1$ in Sect.~\ref{higher}. First, we can shift the parameter $u\to u-u_0\equiv\widetilde{u}$ for some fixed $u_0$ and rewrite $H(u)=T+uV$ as
\beg
\begin{split}
H(u)&=\widetilde{H}(\widetilde{u})\\
&=T(u_0)+\widetilde{u}\,V,
\end{split}
\en
where $T(u_0)=T+u_0V$. The relation between the new $T$-part and $V$ must have the same form as \eref{s}, i.e.
\beg
T(u_0)=W(u_0)+[V, S(u_0)],\quad [V,W(u_0)]=0.
\en
In the present case  $S(u_0)=S$, $W(u_0)=W+u_0V$. For type-1 matrices in particular \eref{cauchy} only changes by a simple $W_E \to W_E+u_0E$.

We can also redefine the parameter as $x=1/u$ and (via multiplication by $x$) transfer the parameter dependence from  $V$ to $T$ and then shift the new parameter $x\to x-x_0\equiv \widetilde{x}$
\beg
\begin{split}
H(x)&=xT +V\\
&=\widetilde{H}(\widetilde{x})\\
&=\widetilde{x}\,T+H(x_0),
\end{split}
\en
where $H(x_0)=x_0T+V$ becomes the new $V$-part. This transformation is more interesting, and has consequences for our construction of type $M>1$ matrices. 

Note that there is an asymmetry in transformation properties under shifts in $u$ and $x$ introduced by our choice to express $T$ through $V$ in \eref{s} rather than the other way around. We have
 \beg
\begin{split}
T=W(x_0) + [H(x_0),S(x_0)],\\  
[H(x_0),W(x_0)]=0.
\end{split}
\label{xTShift}
\en
The $x_0$-dependencies of $W(x_0)$ and $S(x_0)$ are nontrivial. We see that the matrix $T$, and by extension the whole commuting family, is characterized by a continuum of 
antihermitian matrices $S(x_0)$, corresponding to the shift freedom in $x_0$. In particular $S(0)=S$, the unshifted antihermitian matrix.

Specializing to type-1, we understand $S(x_0)$ better by examining the shifted reduced Hamiltonian 
\beg
\begin{split}
\Lambda(x)&=x\ket{\gamma}\bra{\gamma}+E\\
&=\widetilde{\Lambda}(\widetilde{x})\\
&=\widetilde{x}\,\ket{\gamma}\bra{\gamma} +\Lambda(x_0),
\end{split}
\en
from which \eref{xTShift} becomes
\beg
\begin{split}
\ket{\gamma}\bra{\gamma} =W_{\Lambda}(x_0) + [\Lambda(x_0), S(x_0)],\\
[\Lambda(x_0), W_{\Lambda}(x_0)]=0.
\end{split}
\label{xTShiftL}
\en
As in \eref{cauchy}, \eref{xTShiftL} is the defining equation for $S(x_0)$, whose matrix elements obtain most conveniently from the eigenbasis of $\Lambda(x_0)$. 

The matrix $\Lambda(x_0)=x_0\ket{\gamma}\bra{\gamma}+E$ takes the role of $E$  in \eref{cauchy}. In particular,
\beg
S_{ij}(x_0)=\frac{\alpha_i\alpha_j^*}{\lam_i-\lam_j},
\en
where $\lam_i$ are the eigenvalues of $\Lambda(x_0)$ given by \eref{lambdaEq} with $u\to1/x_0$, and $\alpha_i$ are the components of $\ket{\gamma}$ in the eigenbasis of $\Lambda(x_0)$.

\section{Higher types}
\label{higher}

Integrable matrices $H(u) = T+uV$ of type $M \ge 1$ have exactly $n=N-M$ nontrivial linearly independent commuting partners for all $u$. The restriction on $n$ for higher types tends to complicate their parametrizations -- most notably the matrix $V$ is no longer arbitrary. Previous work\cite{owusu1} developed a parametrization (in the eigenbasis of $V$) called the ``ansatz type-$M$'' construction, valid for all $M\ge1$. This construction  is complete for $M=1, 2$ in the sense that one can fit any such integrable matrix into the ansatz construction. Numerical work and parameter counting suggest that it is similarly complete for $M=3$,   but produces   only a subset of measure zero among all type $M>3$ matrices. Finally,  the type-1 construction  of Sect.~\ref{type1r} maps into the ansatz type-1 construction and vice versa.
The parametrization  of Ref.~\onlinecite{owusu1} reads
\beg
\begin{split}
&H_{ij}(u) = T_{ij} = \gamma_i \gamma_j \frac{d_i-d_j}{\eps_i-\eps_j}\frac{\Gamma_i+\Gamma_j}{2},\quad i\ne j,\\
&H_{ii}(u)=uV_{ii}+T_{ii}\\
&= u\,d_i-\sum_{j\ne i}\gamma_j^{2}\frac{d_i-d_j}{\eps_i-\eps_j}\frac{\Gamma_i+\Gamma_j}{2}\frac{\Gamma_j+1}{\Gamma_i+1},
\end{split}
\label{HcompAn1}
\en
where the $\gamma_i$ and $\eps_i$ are free real parameters, and the \textit{constrained} $d_i$ and $\Gamma_i$ obey the following equations with free parameters $g_i$, $P_i$ and $x_0$
\beg
\begin{split}
d_i=\frac{1}{x_0}\sum_{j=1}^{N-M}\frac{g_j}{\braket{j|j}}\frac{1}{\lambda_j-\eps_i},\\
\Gamma_i^2=1+\frac{1}{x_0}\sum_{j=N-M+1}^{N}\frac{P_j}{\braket{j|j}}\frac{1}{\lambda_j-\eps_i},
\end{split}
\label{dGam1}
\en
where $\lambda_i$ and $\braket{i|i}$ are related to $\eps_i$ and $\gamma_i$ through
\beg
\begin{split}
\frac{1}{x_0}=\sum_{j=1}^N\frac{\gamma_j^2}{\lambda_i-\eps_j},\\
\braket{i|i}=\sum_{j=1}^N\frac{\gamma_j^2}{(\lambda_i-\eps_j)^2}.
\end{split}
\label{norms1}
\en
Note that $\lam_i$ and $\ket{i}$ are the eigenvalues and eigenstates, respectively, of a certain \textit{auxiliary} type-1 family, see \esref{lambdaEq} and \re{lambdaEq2}.

The signs of $\Gamma_i$ are arbitrary\cite{Minus1} and each set of sign choices corresponds to a different commuting family. The choice of $x_0$, $\eps_i$ (equivalently $\lambda_i$), $\gamma_i$, and $P_i$\cite{ansatzreal} defines the commuting family while varying $g_i$ produces different matrices within a given family.  Ref.~\onlinecite{owusu1} proves that these equations indeed produce type-$M$ integrable matrices and also determines the eigenvalues of $H(u)$.

\subsection{Rotationally invariant construction}
\label{SecAnsatzRotInv}

Here we present a rotationally invariant formulation of the real symmetric ansatz construction of an $N\times N$ Hamiltonian $H(u)$. We emphasize that unlike the type-1 case we do not have a clear constructive way of motivating the final expressions other than the fact that they reproduce the above basis-specific expressions.

We start with \eref{HcompAn1}.  Consider three mutually commuting real symmetric matrices $V$, $E$ and $\Gamma$. In their shared eigenbasis
\beg
\begin{split}
V&=\textrm{diag}(d_1,d_2,\dots,d_N),\\
E&=\textrm{diag}(\eps_1,\eps_2,\dots,\eps_N),\\
\Gamma&=\textrm{diag}(\Gamma_1,\Gamma_2,\dots,\Gamma_N),\\
 \ket{\gamma}&=(\gamma_1,\gamma_2,\dots,\gamma_N).
\end{split}
\label{commonDiag}
\en
Further, define an antisymmetric matrix $S_M$ through
\beg
\begin{split}
 W_E+[E,S_M]=\frac{\Gamma\ket{\gamma}\bra{\gamma}+\ket{\gamma}\bra{\gamma}\Gamma}{2},\\ 
 [E,W_E]=0.\\
\end{split}
\label{ansatzSdef}
\en
The matrix $T$ obeys
\beg
\begin{split}
T=W_V+[V,S_M], \quad [V,W_V]=0,
\end{split}
\label{ansatzTdef}
\en
which is \eref{s22} with $S\to S_M$. We then require that $(\Gamma+\mathbb{1})\ket{\gamma}$ be an eigenstate of $T$
\beg
 T (\Gamma+\mathbb{1})\ket{\gamma}=0,
\label{ansatzTdef2}
\en
where we set the corresponding eigenvalue to zero via a shift of $T$ by a multiple of the identity. This equation replaces the type-1 equation $T\ket{\gamma}=t\ket{\gamma}$. Basis-independent  Eqs.~(\ref{ansatzSdef}-\ref{ansatzTdef2}) are equivalent to \eref{HcompAn1}.

The next step is to express the constraints \re{dGam1} in a basis-independent form.  To this end we introduce an auxiliary type-1  family with the reduced Hamiltonian  
\beg
\Lambda=x_0\ket{\gamma}\bra{\gamma}+E,
\label{reducedAux}
\en
where we have elected to  transfer the parameter dependence to the $T$-part as discussed in Sect.~\ref{shifting}. We consider this family at a fixed value of the parameter $x=x_0$, so we suppress the dependence on $x_0$ in the reduced Hamiltonian, $\Lambda(x_0)\to\Lambda$, as well as in other members of the auxiliary type-1 family.

By construction $d_i$  are the eigenvalues of  $V$ and   $\Gamma_i^2-1$ are the eigenvalues of a matrix $\Gamma^2-\mathbb{1}$ simultaneously diagonal with $V$.  Multiplying both sides of \eref{dGam1} by $\gamma_i$  and using \esref{lambdaEq} and \re{lambdaEq2}, we see that \eref{dGam1} is equivalent to the following basis-independent equations
\beg
\begin{split}
V\ket{\gamma}=\frac{1}{x_0}\sum_{j=1}^{N-M}\frac{g_j}{\braket{j|j}}\ket{j},\\
(\Gamma^2-\mathbb{1})\ket{\gamma}=\frac{1}{x_0}\sum_{j=N-M+1}^{N}\frac{P_j}{\braket{j|j}}\ket{j}.
\end{split}
\label{dGam3}
\en

It remains to trace parameters $g_i$ and $P_i$ to an object with known transformation properties under a change of basis. By construction, the matrices $V$ and $\Gamma^2-\mathbb{1}$ are simultaneously diagonal with  $V$-parts of the auxiliary type-1 family. We can therefore complement them to the corresponding members of this family as follows
 \beg
 H_1=x_0 T_V+V,\quad H_2=x_0 T_{\Gamma}+\Gamma^2-\mathbb{1},
\label{complement}
\en
 where $T_V$ and $T_{\Gamma}$ are given by \eref{relationsL}. In particular,  $T_V\ket{\gamma}=T_{\Gamma}\ket{\gamma}=0$, so that \eref{dGam3} implies
\beg
\begin{split}
 H_1\ket{\gamma}&=\frac{1}{x_0}\sum_{j=1}^{N-M}\frac{g_j}{\braket{j|j}}\ket{j},\\
 H_2\ket{\gamma}&=\frac{1}{x_0}\sum_{j=N-M+1}^{N}\frac{P_j}{\braket{j|j}}\ket{j}.
\end{split}
\label{complement2}
\en
 Further, since $\ket{j}$ are eigenvectors of $ H_{1,2}$, upon multiplying each side of \eref{complement2} by $\ket{i}\bra{i}$ we find
\beg
\begin{split}
 &H_1\ket{i}=g_i\ket{i}, \quad H_2\ket{i}=0, \quad 1\le i\le N-M,\\
 &H_1\ket{i}=0,\quad  H_2\ket{i}=P_i\ket{i},\quad N-M< i\le N,\\
 \end{split}
\label{complement3}
\en
where we used $\braket{\gamma|j}=x_0^{-1}$, which follows from \esref{lambdaEq} and \re{lambdaEq2}.
Finally, \eref{complement3} implies
\beg
\begin{split}
 H_1 H_2=0.
\end{split}
\label{complement4}
\en
Define $G\equiv H_1+H_2$ to be a real symmetric matrix with $N$ unconstrained eigenvalues $(g_1,g_2,\dots,g_{N-M},P_{N-M+1},\dots,P_N)$. In order to  guarantee that $H(u)$ be real symmetric, however, the numbers $P_j$ and therefore the matrix $G$ must be properly scaled so that the right hand side of the second relation in \eref{dGam1} is nonnegative\cite{ansatzreal}.  

We have therefore derived a basis-independent formulation of Eqs.~(\ref{HcompAn1}-\ref{norms1}) in terms of unconstrained (apart from the aforementioned scaling of $G$ to ensure real $\Gamma$) quantities $(G,E,\ket{\gamma},x_0)$. One works backwards from \eref{complement4} to \eref{ansatzSdef} to derive $(\Lambda,V,\Gamma,T)$ in order to construct ansatz type-$M$ matrices $H(u)=T+uV$. In fact, since \eref{reducedAux} and \eref{complement} imply $[G,\Lambda]=0$, we find it more natural to select $(\Lambda,G,\ket{\gamma},x_0)$ and from them derive $(E,V,\Gamma,T)$. We have no definitive argument, however, that favors one procedure over the other.

Let us now briefly recount the construction. Any real symmetric matrix $G$ allows us to define two matrices $H_1$ and $H_2$ that satisfy \eref{complement4}
\beg
\begin{split}
G=H_1+H_2,\\
H_1H_2=0,
\end{split}
\label{backwards}
\en
where the type $M=\textrm{rank}(H_2)$, the number of non-zero eigenvalues of $H_2$. Let $\Lambda$ be a real symmetric matrix satisfying $[G,\Lambda]=0$. We derive $E$ from $\Lambda$ using \eref{reducedAux}, which generates an auxiliary type-1 integrable family of which $\Lambda$ is the reduced Hamiltonian. Specifically, we obtain the type-1 antisymmetric matrix $S$ through \eref{cauchy}. The common eigenvectors $\ket{i}$ of $\Lambda$, $H_1$ and $H_2$ are given by \eref{lambdaEq2} in the eigenbasis of $E$.

The next step is to obtain $V$ and $\Gamma^2$ through \eref{complement}. To do this we need matrices $T_V$ and $T_{\Gamma}$, for which it is helpful to use the second parameter shift discussed in Sect.~\ref{shifting}. We define the $x_0$-dependent type-1 antisymmetric matrix $S(x_0)$ through \eref{xTShiftL}. Then $T_V$ and $T_{\Gamma}$ are obtained from
\beg
\begin{split}
&T_V=W_1(x_0)+[H_1,S(x_0)],\quad [H_1,W_1(x_0)]=0,\\
&T_{\Gamma}=W_2(x_0)+[H_2,S(x_0)],\quad [H_2,W_2(x_0)]=0,\\
&T_{V,\Gamma}\ket{\gamma}=0,
\end{split}
\label{backwards3}
\en
which when combined with \eref{complement} determines $V$ and $\Gamma^2$. The final step is to determine ansatz $T$ through Eqs.~(\ref{ansatzSdef}-\ref{ansatzTdef2}). The choice of $x_0$, $\ket{\gamma}$, $\Lambda$ and $H_2$ defines the ansatz type-$M$ commuting family, while the choice of $H_1$ specifies a matrix within the family.

Setting $x_0=0$ seemingly simplifies the construction, because then we have $V=H_1$ and $\Gamma^2-1=H_2$ and we bypass the auxiliary type-1 step in the derivation. Despite this simplication, $x_0=0$ actually produces \textit{type-1} integrable matrices $H(u)=T+uV$ with $M$-fold degenerate $V$, which we prove in Appendix~\ref{appendix2}. In this sense, ansatz type-$M$ matrices $H(u)=T+uV$, for which $V$ is generally non-degenerate, are deformations of degenerate type-1 families with deformation parameter $x_0$.

\subsection{Probability distribution function for ensembles of type-$M>1$ integrable matrices}
\label{PDFM}

Despite being significantly more complex than type-1 matrices, ansatz type-$M$ matrices are similarly generated by the choice of two commuting random matrices $G$ and $\Lambda$ and a random vector $\ket{\gamma}$. Therefore, the derivation for the probability density function from Sect.~\ref{PDF1}, restricted to the GOE, also applies to ansatz matrices. Let $c_i$, $1\le i \le N$ be the $N$ eigenvalues of $G$ and $\lambda_i$ those of $\Lambda$. Using \eref{RotInvFinal}
\beg
\begin{split}
P_a(\gamma,c,\lambda) &\propto \delta \left(1- |\gamma|^2\right)\times\\
&\prod_{i<j}|c_i-c_j||\lambda_i-\lambda_j|e^{-\sum_k c_k^2}e^{-\sum_k \lambda_k^2}\\
&=  \delta \left(1- |\gamma|^2\right)P(c)P(\lambda),
\end{split}
\label{RotInvM}
\en
where $(c_1,\dots,c_N)=(g_1,\dots,g_{N-M},P_{N-M+1},\dots,P_N)$ in order to connect \eref{RotInvM} to parameters appearing in  Eqs.~(\ref{HcompAn1}-\ref{norms1}). As noted earlier, one may adopt the alternative viewpoint of selecting the matrix pair $(G,E)$ instead of $(G,\Lambda)$, where there is no commutation restriction on $G$ and $E$. The PDF from this standpoint is then
\beg
\begin{split}
P_b(\gamma,c,\eps) &\propto \delta \left(1- |\gamma|^2\right)\times\\
&\prod_{i<j}|c_i-c_j||\eps_i-\eps_j|e^{-\sum_k c_k^2}e^{-\sum_k \eps_k^2}\\
&=  \delta \left(1- |\gamma|^2\right)P(c)P(\eps),
\end{split}
\label{RotInvM2}
\en
where $\eps_i$ are the eigenvalues of $E$. To be clear, \eref{RotInvM2} and \eref{RotInvM} are two different PDFs for ansatz matrix parameters. To see this, we use \eref{RotInvM} to write down the corresponding $P_a(\gamma,c,\eps)$.
\beg
\begin{split}
P_a(\gamma,c,\eps) &=  \delta \left(1- |\gamma|^2\right)P(c)P(\lambda(\eps,\gamma))\left|\det\frac{\partial \lambda(\eps,\gamma)}{\partial \eps}\right|.
\end{split}
\label{RotInvM3}
\en
There is no \textit{a priori} reason to expect the additional dependence on $\ket{\gamma}$ to cancel out in \eref{RotInvM3}, much less for the resulting PDF to be equal to \eref{RotInvM2}. It is interesting to note that Ref.~\onlinecite{bogo} shows that if $\eps_i$ are GOE or GUE distributed, then $\lambda_i$ will have the same characteristic level repulsion, though this fact alone is insufficient to prove $P_a(\gamma,c,\eps)=P_b(\gamma,c,\eps)$. We have no objective argument that prefers one distribution to the other, although we view $P_a(\gamma,c,\lambda)$ as the more natural choice due to its closer relationship to the type-1 case.

Lastly, we stress that in order for ansatz matrices $H(u)$ to be real symmetric, the parameters $\Gamma_i$ in \eref{HcompAn1} must be real\cite{ansatzreal}. This requirement in turn places the restriction on a given $G$ that the corresponding $P_i$ must be scaled. Therefore, PDFs \eref{RotInvM} and \eref{RotInvM2} are strictly speaking only correct for complex symmetric $H(u)$ and must be modified for real symmetric $H(u)$. For example, one can write $P^{R}_a(\gamma,c,\lambda)=P_a(\gamma,c,\lambda)\mathcal{I}(\gamma,c,\lambda)$ where $\mathcal{I}(\gamma,c,\lambda)$ is a binary indicator function for the condition $\Gamma_i \in \mathbb{R}$.

\section{Discussion}

We derived two basis-independent constructions of integrable matrices $H(u) = T + uV$ that were previously parametrized in a preferred basis -- that of $V$. All type-1 matrices are constructed from Eqs.~(\ref{cauchy}-\ref{relationsL}), while ansatz type-$M\ge1$ are given by  Eqs.~(\ref{ansatzSdef}-\ref{dGam3}) along with Eqs.~(\ref{complement}-\ref{complement4}). The primary significance in obtaining these basis-independent constructions is that one may now speak of and study random ensembles of integrable matrices in the same way that one studies ensembles of ordinary random matrices in random matrix theory (RMT), for which unitary invariance is a theoretical cornerstone\cite{mehta}.

The two invariant constructions involve choosing a vector $\ket{\gamma}$ and two matrices: $E$ and $V$ such that $[E,V]=0$ for type-1, and $\Lambda$ and $G$ such that $[\Lambda,G]=0$ for ansatz type-$M$. We showed that the eigenvalues of $E$ and $V$ come from independent GUE, GOE or GSE eigenvalue distributions. The eigenvalues of $\Lambda$ and $G$, on the other hand come from independent GOE distributions. This result is significant because Ref.~\onlinecite{scaramazza} shows that correlations between these matrix pairs induce level repulsion in integrable matrices, which generally have Poisson statistics.

It follows from the complete type-1 construction presented in Sect.~\ref{type1r} that if $E$, $V$ and $\ket{\gamma}$ are selected from the GUE, GOE or GSE, then the corresponding integrable family of matrices $H^j(u)$ has the same time-reversal properties that define these three ensembles (the ``3-fold way''\cite{dyson,mehta}) for all $u$, and vice-versa. It is possible (though not yet proved) that a similar statement is true for the natural mathematical and physical generalization of these ensembles, initiated by Altland and Zirnbauer\cite{AZ}, that includes charge conjugation (particle-hole) symmetry considerations as well. This ``10-fold way'' is useful in particular for classifying topological insulators and superconductors\cite{10fold}.

Given the known success of RMT in describing generic (e.g. chaotic) quantum Hamiltonians, one can now also study quantum integrability through the lens of an integrable ensemble theory -- integrable matrix theory (IMT). More specifically, until now quantum integrability was mainly studied through specific models satisfying some loose criteria of integrability, whereas there now exists a new tool based on broad and rigorous definitions to study entire classes of quantum integrable models at once. One immediate use of IMT is the study of level statistics in integrable systems, a work soon to be released\cite{scaramazza} by the authors. Another recent development is the proof that the generalized Gibbs ensemble (GGE)\cite{Jaynes,rigolGGE,revGGE} is the correct density matrix for the long-time averages of observables evolving with type-1 Hamiltonians\cite{EYGGE}. An interesting question is how well the GGE works for type $M>1$ matrices under different scalings of $M$ with $N$. Other possibilities include the characterization of localization\cite{disorder} and the reversibility of unitary dynamics\cite{jperes,jgori2,jpros,jcham} generated by matrices in IMT.

There are two further open problems raised in this work that we have not solved. One is the origin and motivation for the ansatz type-$M$ construction found in Sect.~\ref{higher}, which as it stands is verifiably correct but rather ad-hoc in appearance. There ought to be an intuitive motivation for the construction as is the case for the clear and concise type-1 approach found in Sect.~\ref{type1r}. Another open problem is the complete invariant construction of all type $M>3$ matrices, of which only a subset is covered by the ansatz. The reduced Hamiltonian approach to the type-1 solution has an analogous generalization for type-$M$ which could conceivably cover all such matrices, but the details involve working out the general constraints arising from the restricted linear independence of matrices in type-$M$ families, which are nontrivial.
\begin{acknowledgments}

This work was supported in part by the David and Lucille Packard Foundation. The work at UCSC was supported by the U.S. Department of Energy (DOE), Office of Science, Basic Energy Sciences (BES) under Award \# FG02-06ER46319.

\end{acknowledgments}

\appendix
 \section{Degenerate $E$ implies $u$-independent symmetry in type-1 matrices}
 \label{appendix1}
 
In Sect.~\ref{type1r} we constructed $N\times N$ type-1  families starting from a vector $\ket{\gamma}$ and a matrix $E$. The proof that this construction is exhaustive hinges on $E$  being nondegenerate. We show here that a degenerate $E$   implies   a common $u$-independent symmetry prohibited by our definition of an integrable family\cite{note2,degens}.

Suppose  $E$ has a two-fold degeneracy and consider \eref{cauchy} in the eigenbasis of $E$, so that $E=\textrm{diag}(\eps,\eps,\eps_3,\dots,\eps_N)$. We furthermore pick the degenerate subspace of $E$ that diagonalizes $W_E$. The off-diagonal components of \eref{cauchy} read
\beg
\gamma_i \gamma_j^* = (\eps_i-\eps_j)S_{ij}, \quad i\ne j.
\label{offdiagC}
\en
This in particular implies
 that $\gamma_1 \gamma_2^* = 0$ and $S_{12}$ is arbitrary. Without a loss of generality we let $\gamma_1=0$. 
  
Now we turn our attention to $H(u) = T + uV$, where in this basis $V=\textrm{diag}(d_1,d_2,\dots,d_N)$. Note that by definition of type-1 linear independence, for any integrable family there exists an $H(u)$ such that the matrix $V$ is nondegenerate (this is the typical case, but it suffices that there exist  one such matrix). Looking again at off-diagonal components, through \eref{relationsL} we find
\beg
\begin{split}
H_{ij} &= T_{ij} = (d_i-d_j)S_{ij},\quad i\ne j.
\end{split}
\label{offdiagH}
\en
At this point, we can \textit{almost} see that $H(u)$ is block-diagonal, since any $S_{1j}=0$ for $j \ne 2$. In fact, we can visualize $H(u)$ through the following helpful schematic
\[ H(u) = \left(\begin{array}{cccccc}
\times & \times & 0 & 0 & \dots & 0 \\
\times & \times & \times & \times & \dots & \times \\
0 & \times & \times & \times & \dots & \times \\
0 & \times & \times & \times & \dots & \times \\
\dots& & &\dots& &\dots \\
0 & \times & \times & \times & \dots & \times \\
 \end{array} \right),\]
 where $\times$   represents possibly non-zero matrix elements. To show that $H(u)$ is indeed block-diagonal, we consider the eigenvalue equation
\beg
T\ket{\gamma}=t\ket{\gamma},
\label{eigTA}
\en
 which is true by construction of $\Lambda(u)$. Since $\gamma_1=0$, the first component of \eref{eigTA} combined with \eref{offdiagH} implies
\beg
\sum_{j\ne1}(d_1-d_j)S_{1j}\gamma_j=0,
\label{eigTA2}
\en 
and $S_{1j}=0$ for $j \ne 2$ reduces this to
\beg
(d_1-d_2)S_{12} \gamma_2 =0.
\label{eigTA3}
\en
As $V$ is nondegenerate, \eref{eigTA3} requires either $S_{12}=0$ or $\gamma_2=0$. In the first case,  $H(u)$ is of the form
\[ H(u) = \left(\begin{array}{cccccc}
\times & 0 & 0 & 0 & \dots & 0 \\
0 & \times & \times & \times & \dots & \times \\
0 & \times & \times & \times & \dots & \times \\
0 & \times & \times & \times & \dots & \times \\
\dots& & &\dots& &\dots \\
0 & \times & \times & \times & \dots & \times \\
 \end{array} \right),\]
 while in the  second case  $S_{2j}=0$, $j \ne 1$, from  \eref{offdiagC}  and
 \[ H(u) = \left(\begin{array}{cccccc}
\times & \times & 0 & 0 & \dots & 0 \\
\times & \times & 0 & 0 & \dots & 0 \\
0 & 0 & \times & \times & \dots & \times \\
0 & 0 & \times & \times & \dots & \times \\
\dots& & &\dots& &\dots \\
0 & 0 & \times & \times & \dots & \times \\
 \end{array} \right).\]
 Either way, each member of the family $H(u)$ reduces to two such blocks indicating a $u$-independent symmetry.  For example, $\Omega$ made of two similar blocks that are different multiples of identity commutes with $H(u)$.
  
\section{Ansatz matrices at $x_0=0$ are type-1}
 \label{appendix2}

Here we prove that ansatz type-$M$ matrices $H(u)=T+uV$ become type-1 at $x_0=0$, which is most clearly seen in the eigenbasis of $V$. We first review the construction of ansatz matrices $H(u)$ at $x_0=0$. We then construct a particular type-1 family of matrices $\bar{H}(u)$ through Eqs.~(\ref{cauchy}-\ref{relationsL}) and show that $[H(u),\bar{H}(u)]=0$, $\forall u$.

We first consider ansatz type-$M$ matrices $H(u)=T+uV$. At $x_0=0$, \eref{complement} implies that $V=H_1$ and $\Gamma^2-\mathbb{1}=H_2$, so that\cite{Minus1}
\beg
\begin{split}
V&=\textrm{diag}(d_1,d_2,\dots,d_N),\\
&=\textrm{diag}(g_1,g_2,\dots,g_{N-M},0,\dots,0),\\
\Gamma &= \textrm{diag}(\Gamma_1,\Gamma_2,\dots,\Gamma_N),\\
&=\textrm{diag}(1,1,\dots,1,\pm \sqrt{1+P_{N-M+1}},\dots,\pm\sqrt{1+P_N}),\\
E&= \textrm{diag}(\eps_1,\eps_2,\dots,\eps_N).
\end{split}
\label{degenAnsatz}
\en
We note also that $E=\Lambda$ at $x_0=0$ by \eref{reducedAux}. Recall that $E$ arises in the ansatz construction from an auxiliary type-1 problem, so $E$ is nondegenerate without loss of generality (see Appendix~\ref{appendix1}).

With \eref{degenAnsatz} in mind, we also rewrite Eqs.~(\ref{ansatzSdef}-\ref{ansatzTdef2}), the defining equations for the ansatz antisymmetric matrix $S_M$ and for ansatz $T$, which are true at any $x_0$
\beg
\begin{split}
T&=W_V+[V,S_M],\quad [V,W_V]=0,\\
T&\frac{1}{2}(\Gamma+\mathbb{1})\ket{\gamma}=0,
\end{split}
\label{ansatz2}
\en
where $S_M$ follows from
\beg
\begin{split}
\Omega_E+[E,S_M]=\frac{\Gamma\ket{\gamma}\bra{\gamma}+\ket{\gamma}\bra{\gamma}\Gamma}{2},\quad [E,\Omega_E]=0.
\end{split}
\label{SM2}
\en

We now prove that ansatz type-$M$ $H(u)=T+uV$ constructed with \eref{degenAnsatz} are in fact type-1 matrices. Consider a type-1 integrable matrix $\bar{H}(u)=\bar{T}+u\bar{V}$ family constructed through the methods of Sect.~\ref{type1r}, with the substitution $\ket{\gamma}\to\frac{1}{2}(\Gamma+\mathbb{1})\ket{\gamma}\equiv \ket{\bar{\gamma}}$. This particular type-1 family is unrelated to the \textit{auxiliary} type-1 family appearing in the ansatz construction. In the following, bars $\bar{X}$ will indicate quantities $X$ that involve the type-1 integrable matrix family. We have
\beg
\begin{split}
&\bar{V}=\textrm{diag}(\bar{d}_1,\bar{d}_2,\dots,\bar{d}_N)\\
&\ket{\bar{\gamma}}\bra{\bar{\gamma}} = \bar{W}_{E}+[E,\bar{S}],\quad [E,\bar{W}_{E}]=0,\\
&\bar{T} = \bar{W}_{\bar{V}}+[\bar{V},\bar{S}],\quad [\bar{V},\bar{W}_{\bar{V}}]=0,\\
&\bar{T}\ket{\bar{\gamma}}=0,
\label{ansatz3}
\end{split}
\en
where $E$ is the same as in \eref{SM2}, and therefore $[E,\bar{V}]=0$. In particular, the reduced Hamiltonian $\bar{\Lambda}(u)$ (see \eref{introLam}) of this type-1 family is
\beg
\begin{split}
\bar{\Lambda}(u) &=\ket{\bar{\gamma}}\bra{\bar{\gamma}}+uE.
\label{ansatz4}
\end{split}
\en
Recall that by construction $[\bar{\Lambda}(u),\bar{H}(u)]=0$, $\forall u$. Therefore, it suffices to show $[\bar{\Lambda}(u),H(u)]=0$, $\forall u$, which combined with the non-degeneracy of ${\bar\Lambda(u)}$ implies $[\bar{H}(u),H(u)]=0$, $\forall u$.

To this end, consider the commutator $[\bar{\Lambda}(u),H(u)]$
\beg
\begin{split}
&[\bar{\Lambda}(u),H(u)]=\\
&=[\ket{\bar{\gamma}}\bra{\bar{\gamma}},T]+u\left([E,T]+[\ket{\bar{\gamma}}\bra{\bar{\gamma}},V]\right)+u^2[E,V].
\label{x0Comm}
\end{split}
\en
The first term in \eref{x0Comm} vanishes by \eref{ansatz2}, and the third term in \eref{x0Comm} vanishes by construction. We then have
\beg
\begin{split}
&[\bar{\Lambda}(u),H(u)]=u\left([E,T]+[\ket{\bar{\gamma}}\bra{\bar{\gamma}},V]\right).
\label{x0Comm2}
\end{split}
\en
\eref{x0Comm2} is true for all $x_0$, but in order for its r.h.s. to vanish, we must have (see Eqs.~(\ref{relations}-\ref{s}))
\beg
\begin{split}
T=\Omega_V+[V,s],\quad [V,\Omega_V]=0,\\
\ket{\bar{\gamma}}\bra{\bar{\gamma}} = \bar{\Omega}_E+[E,s],\quad [E,\Omega_E]=0,
\label{x0Comm3}
\end{split}
\en
where $s$ is an antisymmetric matrix. \eref{x0Comm3} is not true for general $x_0$, but we can show it is true at $x_0=0$. From \eref{ansatz2}  and \eref{ansatz3} we actually have
\beg
\begin{split}
T = W_V+[V,S_M],\quad [V,W_V]=0,\\
\ket{\bar{\gamma}}\bra{\bar{\gamma}} = \bar{W}_E+[E,\bar{S}], \quad [E,\bar{W}_{E}]=0.
\label{x0Comm4}
\end{split}
\en
We now show that at $x_0=0$, $[V,S_M]=[V,\bar{S}]$, so that $s=\bar{S}$ in \eref{x0Comm3}. This last step will complete the proof that $[H(u),\bar{H}(u)]=0$. Consider the matrix elements $S_{M,ij}$ and $\bar{S}_{ij}$ in the eigenbasis of $V$, which obtain from \eref{SM2} and \eref{ansatz3}
\beg
\begin{split}
S_{M,ij}&= \frac{\gamma_i(\Gamma_i+1)\gamma_j(\Gamma_j+1)}{4}\frac{1}{\eps_i-\eps_j}\\
&-\frac{\gamma_i(\Gamma_i-1)\gamma_j(\Gamma_j-1))}{4}\frac{1}{\eps_i-\eps_j},\\
\bar{S}_{ij}&= \frac{\gamma_i(\Gamma_i+1)\gamma_j(\Gamma_j+1)}{4}\frac{1}{\eps_i-\eps_j},\\
\label{Selements}
\end{split}
\en
but at $x_0=0$, \eref{degenAnsatz} is true and therefore many $\Gamma_i=1$. More precisely, we find
\beg
\begin{split}
S_{M,ij}&= \bar{S}_{ij},\quad \textrm{if }i\le N-M,\textrm{ OR }j\le N-M,\\
S_{M,ij}&\ne \bar{S}_{ij},\quad \textrm{otherwise}.
\label{Selements2}
\end{split}
\en
Now using \eref{degenAnsatz} again, we see that $d_i-d_j=0$ if $S_{M,ij} \ne \bar{S}_{ij}$, where $d_i$ is the $i$-th diagonal entry of the diagonal matrix $V$. Therefore $[V,S_M]=[V,\bar{S}]$ at $x_0=0$, which implies \eref{x0Comm3} holds with $s=\bar{S}$, and therefore $[\bar{\Lambda}(u),H(u)]=0$, $\forall u$. It follows that $[\bar{H}(u),H(u)]=0$, $\forall u$ and $H(u)$ is type-1 at $x_0=0$.


\begin{thebibliography}{99}
\bibitem{note1} Excluding multiples of the identity matrix. Equivalently, one can take $H(u)$ and $H^i(u)$ to be traceless and later add $(c_1+c_2u)\mathbb{1}$ to the family.
\bibitem{note2} A $u$-independent symmetry means  $H^i(u)$ are simultaneously block-diagonal thus reducing the problem to that for smaller matrices.
\bibitem{degens}$N\times N$ Hermitian quaternion-real matrices in general have $N$ distinct real scalar quaternion eigenvalues. Such matrices can also be represented as $2N\times 2N$ complex Hermitian matrices with $N$ distinct real eigenvalues that each appear twice in the spectrum. We clarify that by considering non-degenerate quaternion $E$, we forbid degeneracies in the $N$ quaternion eigenvalues.
\bibitem{rankone} Matrices of this form, i.e. rank-one perturbations to Hermitian matrices have  been previously studied in the context of simple perturbations about non-interacting or mean-field systems, see 
Refs.~\onlinecite{aleiner,bogo} and references therein.
\bibitem{Minus1} We impose $\Gamma_i \ne -1$, for if some $\Gamma_i=-1$, it can be shown that additional constraints arise between ordinarily independent ansatz parameters.
\bibitem{ansatzreal} For unrestricted $P_i$ the matrix $H(u)$ is generally complex symmetric.  However, commutation relations $[H^a(u),H^b(u)]=0$ still hold in this case. In fact, all parameters in our integrable matrix constructions can be unrestricted complex numbers, which result in complex symmetric integrable $H(u)$.
\end{thebibliography}
\end{document}